\documentclass[3p,times,final,fleqn]{elsarticle}

\usepackage{amssymb}
\usepackage{latexsym}

\usepackage{bm}
\usepackage{hyperref}
\usepackage[pagewise,modulo]{lineno}

\journal{ }

\begin{document}


\begin{frontmatter}

\title{Phase transition in the diffusion and bootstrap percolation models on regular random and Erd\H{o}s-R\'{e}nyi networks}

\author[1]{Jeong-Ok Choi}
\author[2]{Unjong Yu\corref{cor1}}
\cortext[cor1]{Corresponding author: 
  Tel.: +82-62-715-3629;  
  fax: +82-62-715-2224;}
\ead{uyu@gist.ac.kr}

\address[1]{Division of Liberal Arts and Sciences, 
            Gwangju Institute of Science and Technology,
            Gwangju 61005, South Korea}
\address[2]{Department of Physics and Photon Science,
            Gwangju Institute of Science and Technology, 
            Gwangju 61005, South Korea}


\begin{abstract}
The diffusion and bootstrap percolation models were studied in regular random and Erd\H{o}s-R\'{e}nyi networks using the modified Newman-Ziff algorithms.
We calculated the percolation threshold and the order parameter of the percolation transition (strength of the giant cluster) and its derivatives.
The percolation transitions are classified by the results.
The diffusion percolation with a small $k$ has a double transition, and the bootstrap percolation with $m\geq3$ has the first-order percolation transition.
The diffusion percolation with a large $k$ and the bootstrap percolation with a small $m$ show the second-order percolation transition.
Particularly, third-order percolation transitions were discovered in the bootstrap percolation of $m=2$ in regular random networks.
\end{abstract}


\end{frontmatter}

\section{Introduction}
Percolation is one of the most important models in statistical mechanics~\cite{Stauffer94}.
Recently, there has been increasing interest in correlated percolation models, where the activity (occupation) of a node (site) depends on the states of its neighbors (see Refs.~\cite{Araujo14,Saberi15} and references therein). 
In this paper, we study the diffusion percolation (DP) and bootstrap percolation (BP) models, which belong to the correlated site percolation model, in complex networks. 
In the DP~\cite{Adler88}, each node is activated with probability $p$, and then inactive nodes that have at least $k$ active neighbors are activated recursively until all the inactive nodes have less than $k$ active neighbors;
In the BP~\cite{Chalupa79}, each node is activated with probability $p$ and then nodes that do not have at least $m$ active neighbors are deactivated recursively until all the active nodes have at least $m$ active neighbors.
The two models are closely related and sometimes both are called BP as well.
The DP of $k$ larger than the maximum degree and the BP of $m=0$ are the same as the classical percolation (CP).

The DP and BP have been studied intensively on lattices~
\cite{Kogut81,Branco84,Khan85,Branco86,Adler88,Adler90,Adler91,Chaves95,Medeiros97,Branco99,Gravner12,Choi19,Choi20}.
In the $\Delta$-regular Archimedean lattices and three-dimensional lattices (simple cubic, body-centered cubic, and face-centered cubic lattices), it was shown that the DP and BP have first-order percolation transitions with the percolation thresholds $p_c=1$ and $p_c=0$, respectively, if $m>m_c$ and $k<(\Delta+1-m_c)$, where $m_c=\lfloor (\Delta+1)/2 \rfloor$ (the only exception is the bounce lattice, which has $m_c=\lfloor (\Delta+1)/2 \rfloor+1$)~\cite{Kogut81,Enter87,Adler88,Branco99,Choi19,Choi20}; 
otherwise, the DP and BP have second-order percolation transitions at finite $p_c$ ($0<p_c<1$) with the same critical exponents as the CP~\cite{Choi19,Choi20} (exceptionally, the BP transition with $m = m_c = 6$ in the face-centered cubic lattice is first-order~\cite{Choi20}).

The BP was originally proposed to understand disordered dilute magnetic systems~\cite{Chalupa79}, and later it turned out that many interesting phenomena of complex systems are also closely related to the DP and BP.
Neuronal activity~\cite{Soriano08}, jamming transition~\cite{Gregorio04,Gregorio05}, and opinion formation~\cite{Ramos15} can be modeled by the DP. 
The diffusion of innovations, which is the diffusion process of a new idea, technology, or product through a social network~\cite{Rogers03}, can be modeled by the DP or BP depending on the properties of the innovative option. 
The DP can describe the situation where an agent adopts the innovative option when at least $k$ neighbors have adopted it~\cite{Centola10,Jin21}. 
On the other hand, the BP is a more appropriate model for a situation that an agent abandons the innovative option whenever there are less than $m$ neighbors that keep the innovative option~\cite{Helbing12}.
Recently, there have been studies about the network robustness~\cite{Buldyrev10,Parshani10,Hu11,Gao12,Havlin15}, which adopt the culling process of the BP.
They considered two networks that interact with each other; a failure of a node in one network leads to the failure of some other nodes in the other network. The iterative cascade of failures may make the global connectivity of the two networks broken.
Interestingly, they found a critical point that separates first- and second-order percolation transitions~\cite{Parshani10,Hu11,Gao12}.

Studies of the DP and BP in complex networks, however, are still insufficient and mostly restricted to the DP.
Balogh and Pittel studied the minimum value of $p$ that can make all the nodes active by the DP in regular random networks (RRNs)~\cite{Balogh07}. 
Baxter et al. discovered a double transition (continuous percolation transition followed by a discontinuous hybrid transition) for the DP of $k=3$ in the Erd\H{o}s-R\'{e}nyi network (ERN) with an average degree $\langle\Delta\rangle=5$~\cite{Baxter10}. 
The double transition of the DP was also observed in small-world networks and an RRN~\cite{Gao15}. 
Multiple hybrid phase transition of the DP was reported in ERNs with community structure~\cite{Wu14}.

In this paper, we present comprehensive results on percolation transitions of the DP and BP in RRNs and ERNs using modified Newman-Ziff algorithms. The percolation transitions are classified by the behavior of the order parameter and its derivatives near the percolation threshold;
first-order, second-order, and double transitions are observed. 
Notably, we also discovered third-order percolation transitions for the BP with $m=2$ in RRNs.

\section{Methods}

The strength of the largest cluster ($P_{\infty}$), which is the probability that a node belongs to the largest cluster, is the order parameter of the percolation transition~\cite{Stauffer94}.
We calculated $P_{\infty}$ for the DP and BP using the modified Newman-Ziff algorithms.
In the Newman-Ziff algorithm~\cite{Newman00,Newman01}, the average strength of the largest cluster $P_{\infty}(n)$ is calculated as a function of the number of initially active nodes ($n$), and $P_{\infty}(p)$ as a function of the initial probability of active nodes $p$ is obtained using the following convolution:
\begin{eqnarray}
 P_{\infty}(p) = \sum_{n=0}^{N} B_p(n,N) \;  P_{\infty}(n) , \label{canonical_transf}
\end{eqnarray}
where $B_p(n,N)={_NC_n} \, p^n (1-p)^{N-n}$ is the binomial distribution with ${_NC_n}={N!}/[n! (N-n)!]$. Here $N$ is the total number of nodes. 
Since $P_{\infty}(p)$ for any value of $p$ can be calculated easily if $P_{\infty}(n)$ is obtained once, the Newman-Ziff algorithm is much more efficient than traditional brute-force methods.
Another advantage of the Newman-Ziff algorithm is that derivatives of a physical quantity can be obtained without numerical differentiation, which has inevitably a large error. 
The first, second, and third derivatives of $P_{\infty}(p)$ can be obtained by
\begin{eqnarray}
\frac{d P_{\infty}(p)}{dp} &=& \frac{d}{dp} \left[\sum_{n=0}^{N} B_p(n,N)
                            \, P_{\infty}(n) \right]  = \sum_{n=0}^{N} B_p(n,N) \frac{n-Np}{p(1-p)} \, 
                            P_{\infty}(n)  \\
\frac{d^2 P_{\infty}(p)}{dp^2} &=& \frac{d}{dp} \left[\sum_{n=0}^{N} B_p(n,N) \, \frac{n-Np}{p(1-p)} \, 
    P_{\infty}(n)  \right] = \sum_{n=0}^{N} B_p(n,N) \, \frac{p(N-1)(Np-2n) + n(n-1)}{p^2(1-p)^2} \, 
    P_{\infty}(n) \\
\frac{d^3 P_{\infty}(p)}{dp^3} &=& \frac{d}{dp} 
    \left[ \sum_{n=0}^{N} B_p(n,N) \, \frac{p(N-1)(Np-2n) + n(n-1)}{p^2(1-p)^2} \, 
    P_{\infty}(n) \right] \nonumber \\
&=& \sum_{n=0}^{N} B_p(n,N) \, \frac{p^2(N-1)(N-2)(3n-Np)+n(n-1)[3p(2-N)+n-2]}{p^3(1-p)^3} \,
    P_{\infty}(n) . \label{canonical_diff}
\end{eqnarray}
The derivatives of $P_{\infty}(p)$ are used to determine the percolation threshold and the order of percolation transitions.
Equations~(\ref{canonical_transf})-(\ref{canonical_diff}) are generally applied to other physical quantities.

The Newman-Ziff algorithm was modified for the DP and BP~\cite{Choi19}. 
In the case of the DP, at each step of the Newman-Ziff algorithm, all nodes with at least $k$ active neighbors are activated recursively. 
Pre-active (pre-occupied) state is introduced for the BP: the chosen node at each Newman-Ziff step is not activated but is set to be a pre-active state, which becomes active only when it has at least $m$ active neighbors. 
Using these algorithms, the DP and BP can be studied as efficiently as the CP except for the BP with $m\geq3$. 
As for the BP with $m=3$, it takes about ten times more CPU time than the CP~\cite{Choi19}, and it becomes worse for larger $m$. 
To solve this problem, we proposed another algorithm for the BP in regular networks using the close relation of the DP and BP~\cite{Choi20}.
The algorithm was used for three-dimensional lattices successfully but it is inapplicable to nonregular networks.
In this paper, we extend the algorithm of Ref.~\cite{Choi20} to simulate BP with a large $m$ in nonregular networks.
Within the Newman-Ziff algorithm, the BP process begins from the initial state ${S_n}$ with $n$ active nodes at random. Then any node with less than $m$ active neighbors is deactivated (culled out) recursively.
This culling process of active nodes can be regarded as a diffusion process of inactive nodes with the new rule that active node with less than $m$ active neighbors is deactivated. 
Therefore, the BP process from $S_n$ is equivalent to the diffusion process from $S_{N-n}$ with the new diffusion rule.
Note that nodes with less than $m$ neighbors should be excluded in the BP activation. The new algorithm can be implemented as follows.
\renewcommand{\labelenumi}{(\arabic{enumi})}
\begin{enumerate}
\item Initially, all nodes are set active. \label{item1}
\item Make an array of all the nodes in random order. \label{item2}
\item Set the step number to be $n'=1$.
\item Deactivate the $n'$-th node of the array made in step~(\ref{item2}). Deactivate any node that has less than $m$ active neighbors recursively. 
If the newly deactivated node has at least $m$ neighbors, push the node in the stack with the step number $n'$.
Increase $n'$ by one. \label{item4}
\item  Repeat step~(\ref{item4}) until all the nodes are deactivated.  \label{item5}
\item  Pop all the nodes with $n'$ from the stack and activate them. 
    Calculate $P_{\infty}(n)$ with $n=N-n'$.  \label{item6}
\item  Repeat step~(\ref{item6}) until the stack is empty.
\end{enumerate}
From step~(\ref{item1}) to step~(\ref{item5}), the diffusion process from $S_{N-n}$ with the new diffusion rule is saved in the stack; the stack is a linear data structure in the last-in-first-out (LIFO) order.
The information of the stack is used to simulate the BP process from $S_n$.
The whole steps are repeated to make an average,
and the transformations of Eqs.~(\ref{canonical_transf})-(\ref{canonical_diff}) give $P_{\infty}(p)$ and its derivatives.
This algorithm gives mathematically the same results as that in Ref.~\cite{Choi19}.
For $\Delta$-regular networks, the algorithm of Ref.~\cite{Choi20} is the most efficient since the DP and BP can be simulated simultaneously by one calculation for $m+k=\Delta+1$.
For nonregular networks, on the other hand, the algorithm of Ref.~\cite{Choi19} and the new algorithm should be used separately for the DP and BP, respectively.
The CPU time required for this algorithm is of the same order as the Newman-Ziff algorithm for the original percolation; it is proportional to the network size $N$ and average degree $\langle\Delta\rangle$ as $T_{\mathrm{cpu}} \sim \langle\Delta\rangle^{0.5} N^{1.2}$~\cite{Choi19}. It is notable that this algorithm can be efficiently parallelized within distributed memory techniques such as message passing interface (MPI).

In this work, we consider two kinds of complex networks: RRNs and ERNs.
RRNs with $N$ nodes of degree $\Delta$ are generated by the Steger-Wormald algorithm~\cite{Steger99,Choi20PhA}, which produces uniform RRNs asymptotically in the limit of large $N$~\cite{Kim03}.
ERNs are formed by the standard method~\cite{Erdos59}: arbitrary two nodes are connected by a link with probability $p_{\mathrm{ER}}$ and so the average degree is $\langle\Delta\rangle=(N-1) p_{\mathrm{ER}}$.
We studied RRNs and ERNs with $3\leq\Delta\leq10$ and $3\leq\langle\Delta\rangle\leq10$, respectively.
We made twenty realizations of networks for each parameter set; we confirmed that every realization gives equivalent results within statistical error bars.

\begin{figure}
\includegraphics[width=15.0cm]{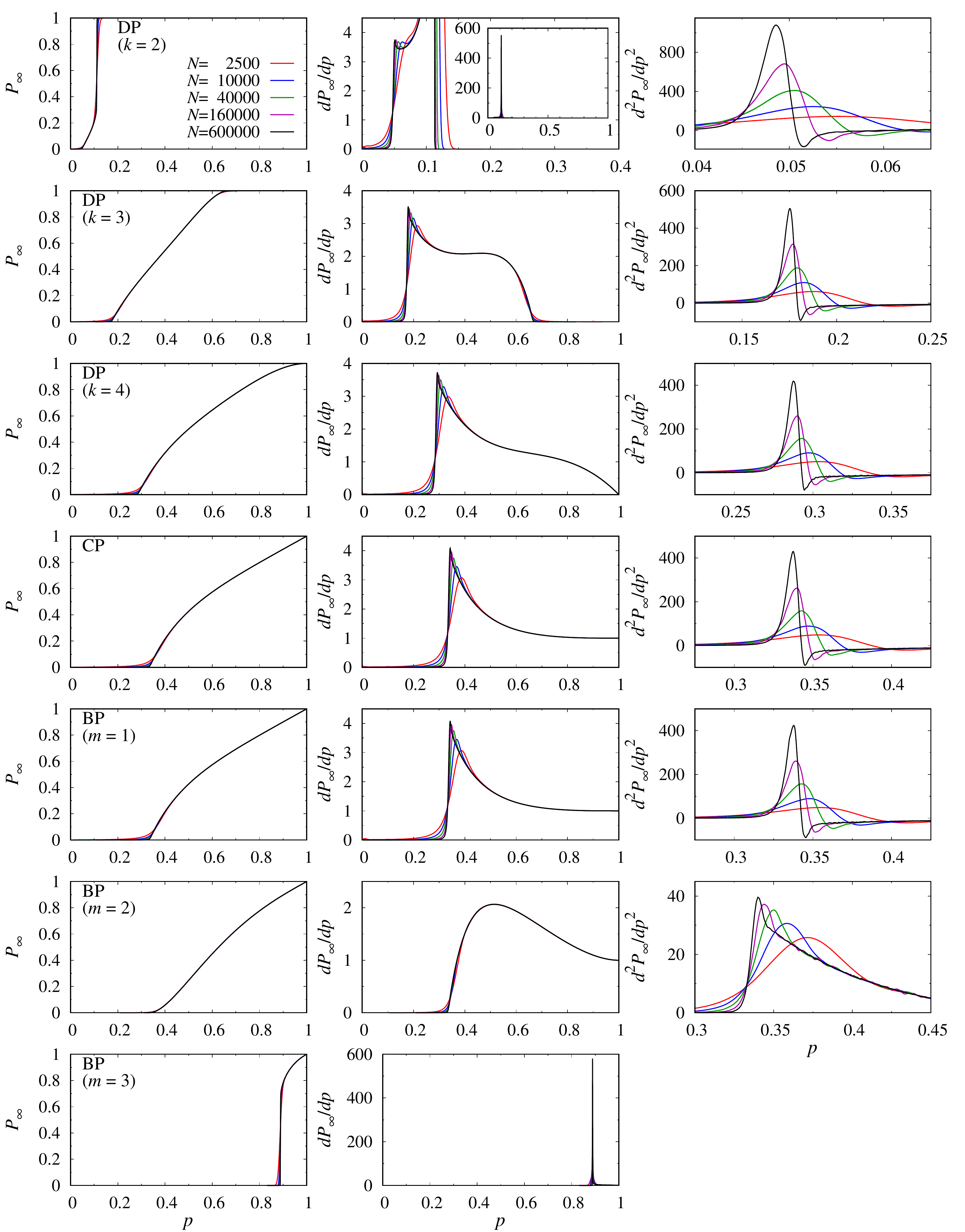}
\caption{Strength of the largest cluster ($P_{\infty}$) and its derivatives as a function of initial filling probability $p$ for the DP, CP, and BP in regular random networks with $\Delta=4$. $N$ represents the number of nodes in the network.}
\label{Fig1}
\end{figure}

\begin{figure}
\includegraphics[width=15.0cm]{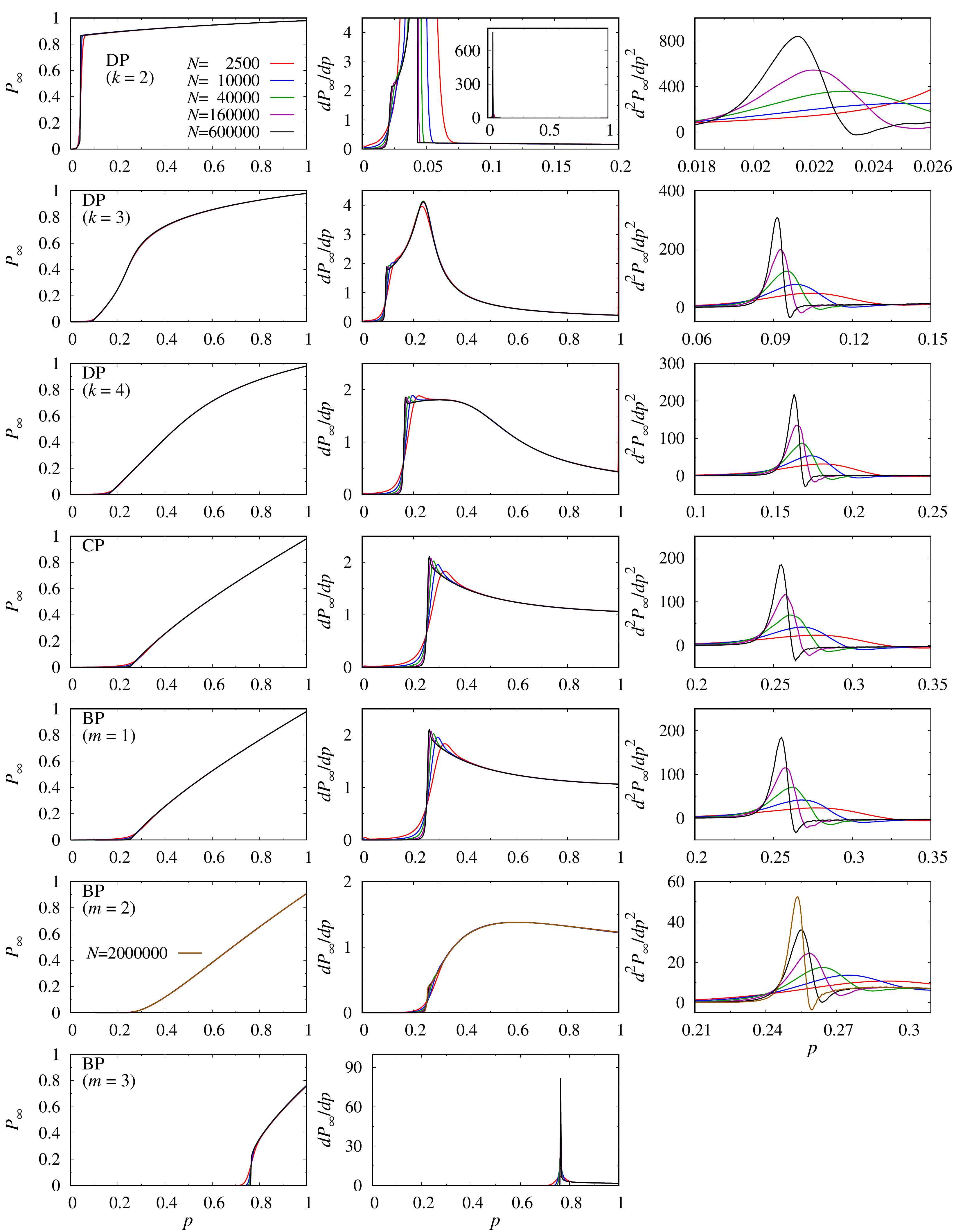}
\caption{Strength of the largest cluster ($P_{\infty}$) and its derivatives as a function of initial filling probability $p$ for the DP, CP, and BP in Erd\H{o}s-R\'{e}nyi networks with $\langle\Delta\rangle=4$.}
\label{Fig2}
\end{figure}

\section{Results and discussions}

Figures~\ref{Fig1} and \ref{Fig2} show strength of the largest cluster ($P_{\infty}$) and its first and second derivatives as a function of initial filling probability $p$ for the DP, CP, and BP in RRN of $\Delta=4$ and ERNs of $\langle\Delta\rangle=4$, respectively. 
For the DP, CP, and BP of $m\leq2$, the percolation transition is continuous, and it is discontinuous (first-order) for the BP of $m=3$.
Note the double transition in the DP with $k=2$:
the second transition, which shows a discontinuous jump of $P_{\infty}$, appears at $p_{c2}$ with $p_{c2}>p_c$. 
The transition at $p_{c2}$ is the hybrid transition, which combines a discontinuity and a singularity~\cite{Baxter10,Lee16}.
The three kinds of transition type (double transition, continuous, and first-order transition) are also observed in RRNs and ERNs of the other average degree.

The phase transition can be classified, in general, by the discontinuity of derivatives of the free energy~\cite{Ehrenfest33,Sauer17}.
The discontinuity can appear in two types. 
In the magnetic phase transition of the two-dimensional Ising model~\cite{Ising25}, for example, the magnetic susceptibility and specific heat (second-order derivatives of the free energy) diverge at the transition temperature~\cite{Onsager44}. 
In the case of the lambda transition~\cite{Keesom32}, to the contrary, there is a jump discontinuity in the specific heat.
If a discontinuity appears in the $n$th-order derivative of the free energy, it is called the $n$th-order phase transition.
If the first derivative of the free energy (e.g., order parameter) has a discontinuity at the transition point,
it is the first-order phase transition; otherwise, the phase transition is classified as the continuous phase transition.
With rare cases such as the Berezinskii-Kosterlitz-Thouless transition~\cite{BKT1,BKT2}, which is an infinite-order phase transition, most of the continuous phase transitions are of second-order. 
Recently, third-order phase transitions were proposed concerning large-$N$ four-dimensional lattice gauge theory~\cite{Gross80}, spin glass~\cite{Crisanti03}, supercritical fluids~\cite{Koga09}, the domino tilings of an Aztec diamond~\cite{Colomo13}, and constrained Coulomb gas~\cite{Cunden17}.
In some superconducting transitions, third-order~\cite{Junod99,Werner99} and fourth-order~\cite{Kumar99,Hall00} phase transitions were suggested.
However, there are still controversies about the existence of higher-order phase transitions~\cite{Woodfield99,Wang11}.
In the case of the percolation transition, the order of the phase transition can be defined by the discontinuity of the order parameter $P_{\infty}$ and its derivatives: 
if $P_{\infty}$ has a discontinuity at the transition, the percolation transition is classified as first-order~\cite{Gao12,Lee16}.
It is well-known that the classical percolation transition is of second-order~\cite{Stauffer94}. 
However, recently, explosive percolation models~\cite{Achlioptas09} show very abrupt continuous or discontinuous phase transition depending on the spatial dimension and detailed percolation rules~\cite{Ziff09,daCosta10,Riordan11,Cho13}. (For a review, see Refs.~\cite{Boccaletti16,Lee16}.)
As for the DP and BP in lattices, the percolation transition is also second-order when the percolation threshold is between 0 and 1~\cite{Choi19,Choi20}. (Only one exception is known: the BP transition with $m=6$ in the face-centered cubic lattice at $p_c=0.75243(6)$ is first-order~\cite{Choi20}.)

\begin{figure}
\includegraphics[width=15.0cm]{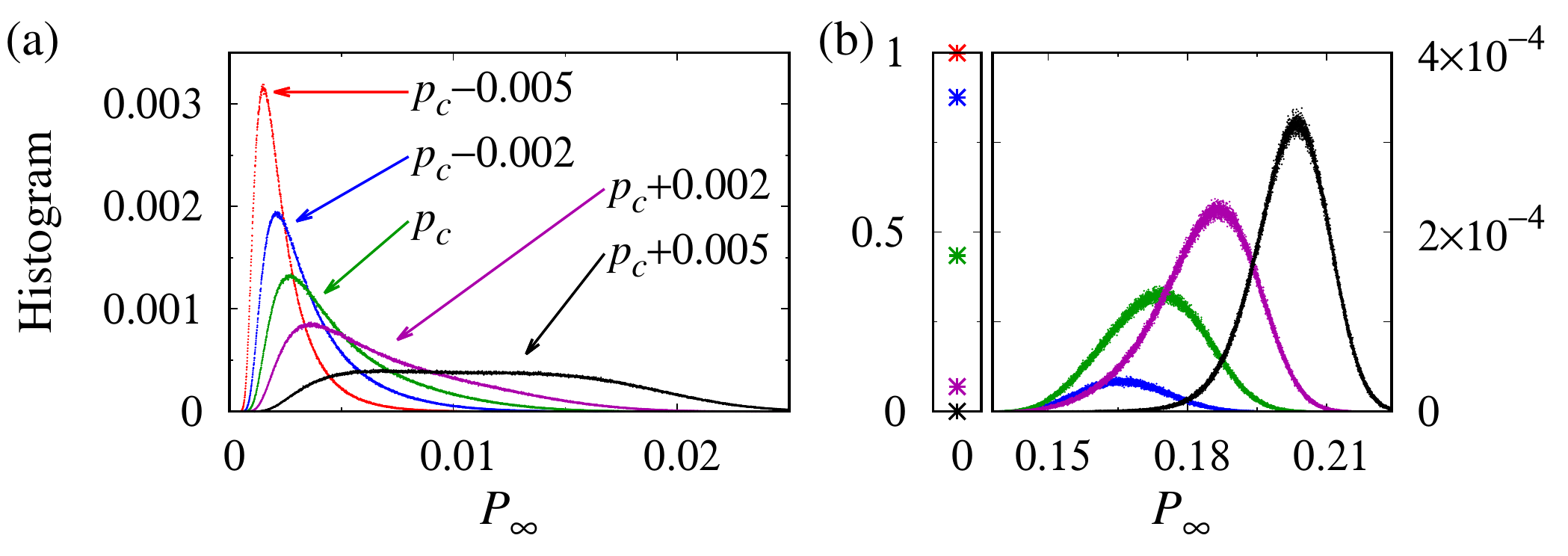}
\caption{Histogram of the strength of the largest cluster ($P_{\infty}$)  for the bootstrap percolation with $m=1$ in (a) and $m=3$ in (b) when initial filling probability $p$ is $p_c\pm 0.005$, $p_c\pm 0.002$, and $p_c$, where $p_c$ is the percolation threshold. The network is regular random with $\Delta=9$ and $N=160\,000$.}
\label{Fig3}
\end{figure}

\begin{figure}
\includegraphics[width=15.0cm]{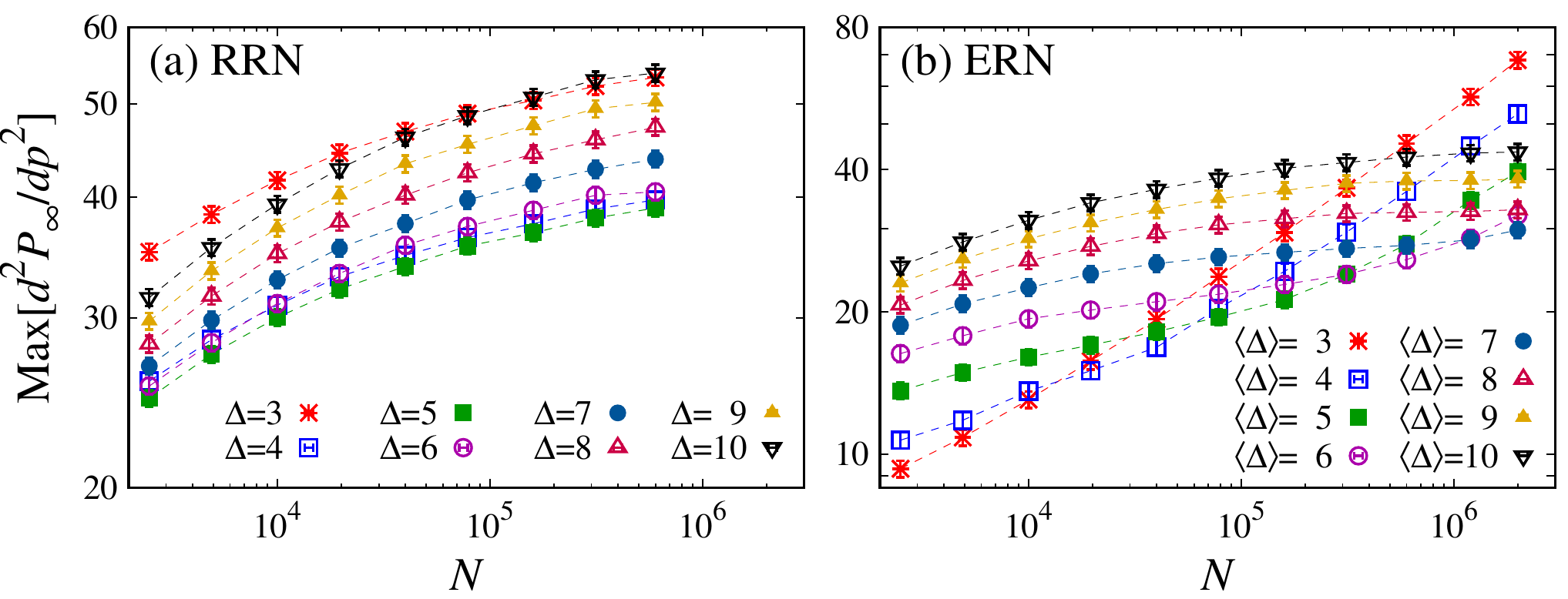}
\caption{The maximum value of the second derivative of the strength of the largest cluster ($d^2P_{\infty}/dp^2$) as a function of the network size ($N$) of the BP  with $m=2$ in regular random networks (RRNs) and Erd\H{o}s-R\'{e}nyi networks (ERNs) in log-log scale. Dashed lines serve as an eye guide.}
\label{Fig4}
\end{figure}

\begin{figure}
\includegraphics[width=15.0cm]{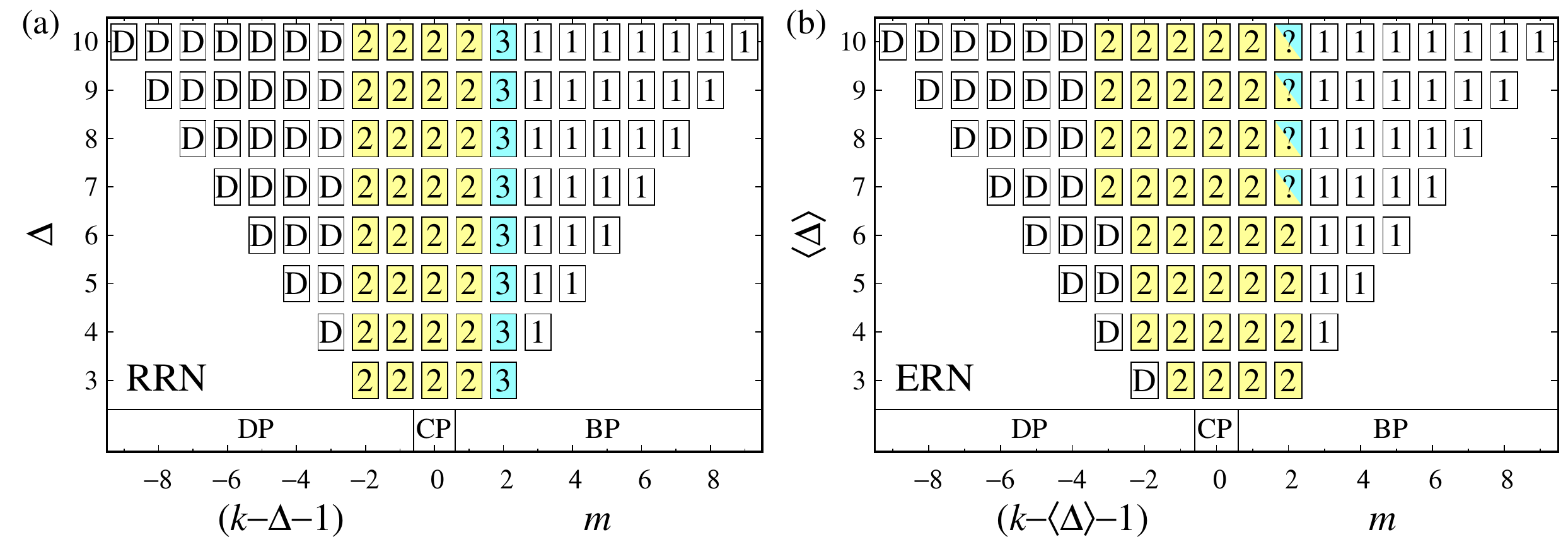}
\setlength{\fboxsep}{1pt}
\caption{Types of the percolation phase transition of the DP, CP, and BP in regular random networks (RRNs) and Erd\H{o}s-R\'{e}nyi networks (ERNs).
\fbox{D}, \fbox{1}, \fbox{2}, and \fbox{3} represent double, first-order, second-order, and third-order transitions, respectively.
\fbox{?} means that it is not clear whether the transition is of second-order or of third-order.}
\label{Fig5}
\end{figure}

As shown in Figs.~\ref{Fig1} and \ref{Fig2}, the DP and BP transitions in RRNs and ERNs are continuous except for the BP of $m\geq 3$, which is first-order. 
Figure~\ref{Fig3} shows the histogram of the order parameter near the percolation threshold, which distinguishes continuous and first-order transitions clearly~\cite{Selke10,Azhari20}.
In the case of a continuous transition (the BP of $m=1$), there is only one peak in the histogram and the peak moves to higher $P_{\infty}$ as $p$ increases. 
Near the first-order transition (the BP of $m=3$), to the contrary, there are two peaks, and the weights of the two peaks change with $p$. 

Note that the BP of $m=2$ in RRNs shows the discontinuity not in $dP_{\infty}/dp$ but in $d^2 P_{\infty}/dp^2$, which means that the percolation transition is of third-order. 
The discontinuity can be clarified by the maximum value of $d^2 P_{\infty}/dp^2$ as a function of the network size. 
For second-order transitions, the maximum value of $d^2 P_{\infty}/dp^2$ diverges as $d^2 P_{\infty}/dp^2 \propto N^{\lambda}$, where $\lambda>0$ is a fitting parameter; this shows the discontinuity of $dP_{\infty}/dp$ at $p=p_c$. 
For the BP of $m=2$ in RRNs, however, the maximum value of $d^2 P_{\infty}/dp^2$ is saturated as shown in Fig.~\ref{Fig4}(a), which implies the absence of discontinuity in $dP_{\infty}/dp$;
we confirmed that the maximum value of $d^3 P_{\infty}/dp^3$ diverges as $N$ increases.
In the case of the BP of $m=2$ in ERNs, as shown in Fig.~\ref{Fig4}(b), $d^2 P_{\infty}/dp^2$ behaves like in RRNs for small networks ($N<N_c$), but for $N>N_c$, it shows a diverging behavior as $N$ increases. The critical size $N_c$ increases as average degree $\langle\Delta\rangle$ increases roughly as $N_c \approx 6\exp(2\langle\Delta\rangle)$, and the divergence was not observed up to $N=2\,000\,000$ for $\langle\Delta\rangle \geq 7$.
Therefore, it would be reasonable to assume that the transition is of second-order in infinite networks irrespective of $\langle\Delta\rangle$.

Figure~\ref{Fig5} summarizes the types of the percolation transitions. 
As for the RRNs, double transition behavior appears for the DP with $k \leq \Delta-2$. 
Second-order transition is for the DP with $k \geq \Delta-1$, for the CP, and for the BP with $m=1$. 
The BP of $m=2$ has a third-order transition and the BP with $m\geq3$ shows the first-order transition. 
In the ERNs, the same sequence of phase transition type is also observed but the existence of a third-order percolation transition is questionable.

\begin{figure}
\includegraphics[width=15.0cm]{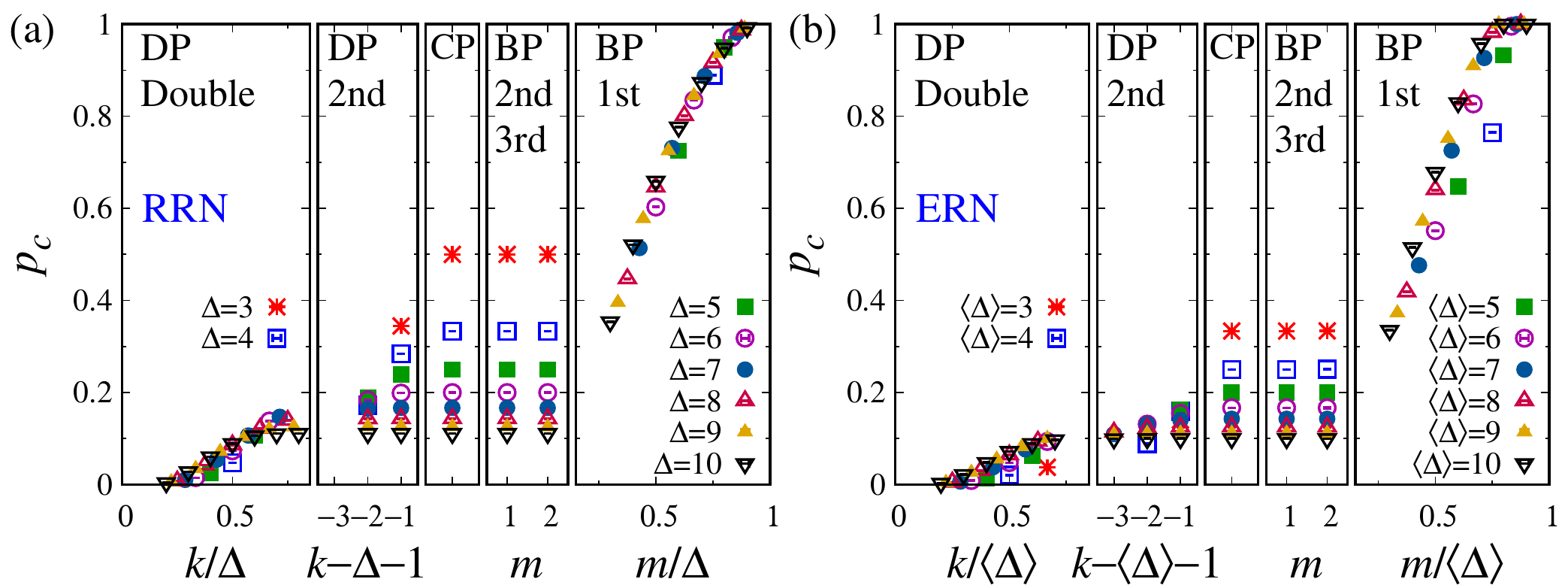}
\caption{Percolation threshold for the DP, CP, and BP in regular random networks (RRNs) and  Erd\H{o}s-R\'{e}nyi networks (ERNs). ``1st'', ``2nd'', and ``3rd'' represent first-, second-, and third-order percolation transitions, respectively.
In the left panels, ``Double'' means second-order percolation followed by a discontinuous jump of $P_{\infty}$.}
\label{Fig6}
\end{figure}

\begin{table}
\caption{\label{tableRR} Percolation threshold ($p_{c}$) as a function of average degree ($\langle\Delta\rangle$) of the diffusion percolation (DP) and bootstrap percolation (BP) in regular random networks (RRNs) and Erd\H{o}s-R\'{e}nyi networks (ERNs).
The second transition probability ($p_{c2}$) is also provided for the double transition.}
\centering
\begin{tabular}{|c|c|cc|cc|c|c|c|}
\hline
  & & \multicolumn{2}{|c|}{RRN-DP} & \multicolumn{2}{|c|}{ERN-DP} & & RRN-BP & ERN-BP\\
 $\langle\Delta\rangle$ & $k$ & $p_{c}$ & $p_{c2}$ & $p_{c}$ & $p_{c2}$ & $m$ & $p_{c}$ & $p_{c}$\\ \hline
3 & $2$ & 0.1250(1) & -         & 0.0374(4) & 0.0940(2) & $1$ & 0.5000(1) & 0.3335(4) \\
  & $3$ & 0.3444(1) & -         & 0.1399(4) & -         & $2$ & 0.5000(1) & 0.3336(4) \\ \hline
4 & $2$ & 0.0472(1) & 0.1111(1) & 0.0201(3) & 0.0408(2) & $1$ & 0.3332(1) & 0.2500(2) \\
  & $3$ & 0.1723(1) & -         & 0.0886(4) & -         & $2$ & 0.3332(1) & 0.2503(4) \\ 
  & $4$ & 0.2840(1) & -         & 0.2840(1) & -         & $3$ & 0.8889(2) & 0.7644(4) \\ \hline
5 & $2$ & 0.0245(1) & 0.0508(2) & 0.0129(1) & 0.0242(1) & $1$ & 0.2499(1) & 0.2000(1) \\
  & $3$ & 0.1060(1) & 0.2751(1) & 0.0630(1) & 0.1366(3) & $2$ & 0.2499(1) & 0.2000(3) \\ 
  & $4$ & 0.1892(1) & -         & 0.1190(1) & -         & $3$ & 0.7249(2) & 0.6475(2) \\ 
  & $5$ & 0.2399(1) & -         & 0.1630(4) & -         & $4$ & 0.9492(2) & 0.9326(2) \\ \hline
6 & $2$ & 0.0150(1) & 0.0291(1) & 0.0089(1) & 0.0161(2) & $1$ & 0.2000(1) & 0.1666(2) \\
  & $3$ & 0.0732(1) & 0.1651(2) & 0.0476(2) & 0.0938(3) & $2$ & 0.1999(1) & 0.1668(5) \\ 
  & $4$ & 0.1380(1) & 0.3972(2) & 0.0935(2) & 0.2298(2) & $3$ & 0.6028(2) & 0.5514(3) \\ 
  & $5$ & 0.1837(1) & -         & 0.1314(2) & -         & $4$ & 0.8349(2) & 0.8267(4) \\ 
  & $6$ & 0.1987(1) & -         & 0.1545(1) & -         & $5$ & 0.9709(1) & 0.9956(4) \\ \hline
7 & $2$ & 0.0101(1) & 0.0188(1) & 0.0065(1) & 0.0115(1) & $1$ & 0.1666(1) & 0.1428(1) \\
  & $3$ & 0.0542(1) & 0.1129(1) & 0.0376(1) & 0.0703(2) & $2$ & 0.1665(1) & 0.1429(2) \\ 
  & $4$ & 0.1067(1) & 0.2690(2) & 0.0763(1) & 0.1668(3) & $3$ & 0.5137(2) & 0.4764(1) \\ 
  & $5$ & 0.1471(1) & 0.4863(2) & 0.1096(2) & -         & $4$ & 0.7310(2) & 0.7255(1) \\ 
  & $6$ & 0.1641(1) & -         & 0.1309(2) & -         & $5$ & 0.8871(2) & 0.9267(2) \\ 
  & $7$ & 0.1665(1) & -         & 0.1398(2) & -         & $6$ & 0.9812(1) & 1.0000(1) \\ \hline
8 & $2$ & 0.00726(1)& 0.0132(1) & 0.00500(4)& 0.0086(1) & $1$ & 0.1429(1) & 0.1250(2) \\
  & $3$ & 0.0422(1) & 0.0832(2) & 0.0308(4) & 0.0555(1) & $2$ & 0.1428(1) & 0.1250(3) \\ 
  & $4$ & 0.0859(1) & 0.1992(2) & 0.0641(2) & 0.1317(2) & $3$ & 0.4468(2) & 0.4181(4) \\ 
  & $5$ & 0.1217(1) & 0.3540(2) & 0.0938(2) & 0.2335(2) & $4$ & 0.6459(2) & 0.6402(2) \\ 
  & $6$ & 0.1391(1) & 0.5532(2) & 0.1135(4) & -         & $5$ & 0.8008(2) & 0.8350(2) \\ 
  & $7$ & 0.1426(1) & -         & 0.1222(2) & -         & $6$ & 0.9168(2) & 0.9816(2) \\ 
  & $8$ & 0.1428(1) & -         & 0.1246(2) & -         & $7$ & 0.9868(1) & 1.0000(1) \\ \hline
9 & $2$ & 0.00548(1)& 0.0098(1) & 0.0039(2) & 0.0067(2) & $1$ & 0.1250(1) & 0.1111(1) \\
  & $3$ & 0.0340(1) & 0.0645(2) & 0.0257(1) & 0.0452(2) & $2$ & 0.1249(1) & 0.1110(2) \\ 
  & $4$ & 0.0713(1) & 0.1557(2) & 0.0548(2) & 0.1081(3) & $3$ & 0.3949(2) & 0.3723(3) \\ 
  & $5$ & 0.1032(1) & 0.2755(2) & 0.0815(3) & 0.1896(4) & $4$ & 0.5771(2) & 0.5713(5) \\ 
  & $6$ & 0.1203(1) & 0.4229(2) & 0.0999(2) & 0.2925(2) & $5$ & 0.7245(2) & 0.7510(6) \\ 
  & $7$ & 0.1245(1) & 0.6051(2) & 0.1083(3) & -         & $6$ & 0.8443(2) & 0.9091(8) \\ 
  & $8$ & 0.1250(1) & -         & 0.1107(1) & -         & $7$ & 0.9355(2) & 1.0000(1) \\ 
  & $9$ & 0.1250(1) & -         & 0.1111(2) & -         & $8$ & 0.9902(1) & 1.0000(1) \\ \hline
10& $2$ & 0.00428(5)& 0.0075(1) & 0.0032(1) & 0.0054(1) & $1$ & 0.1111(1) & 0.1000(2) \\
  & $3$ & 0.0282(1) & 0.0519(1) & 0.0219(1) & 0.0378(2) & $2$ & 0.1110(1) & 0.0998(2) \\ 
  & $4$ & 0.0604(1) & 0.1263(2) & 0.0478(2) & 0.0912(2) & $3$ & 0.3537(2) & 0.3351(3) \\ 
  & $5$ & 0.0891(1) & 0.2234(2) & 0.0720(2) & 0.1597(1) & $4$ & 0.5207(1) & 0.5145(4) \\ 
  & $6$ & 0.1058(1) & 0.3407(2) & 0.0891(2) & 0.2422(1) & $5$ & 0.6593(2) & 0.6783(4) \\ 
  & $7$ & 0.1105(1) & 0.4793(1) & 0.0972(3) & 0.3448(2) & $6$ & 0.7766(2) & 0.8296(3) \\ 
  & $8$ & 0.1111(1) & 0.6463(2) & 0.0995(2) & -         & $7$ & 0.8737(2) & 0.9585(2) \\ 
  & $9$ & 0.1111(1) & -         & 0.0999(2) & -         & $8$ & 0.9481(1) & 1.0000(1) \\ 
  & $10$& 0.1111(1) & -         & 0.1000(2) & -         & $9$ & 0.9925(1) & 1.0000(1) \\  \hline
\end{tabular}
\end{table}

We used two kinds of methods to calculate the percolation threshold $p_c$.
For $n$th-order percolation transitions, the finite-size percolation threshold $p_c(N)$ for a finite network of size $N$ is defined by the initial probability that maximizes $d^n P_{\infty}/dp^n$. 
The percolation threshold of infinite network ($p_c$) can be obtained by the scaling relation $[p_c(N) - p_c] \propto N^{-a}$ with a positive fitting parameter $a$.
The other method uses the fact that values $d^{n-1} P_{\infty}/dp^{n-1}$ of different sizes coincide at $p_c$ for $n$th-order percolation transitions, which is confirmed in Figs.~\ref{Fig1} and \ref{Fig2}.
The two methods give equivalent results within error bars. 
The percolation threshold of the DP, CP, and BP obtained by these methods are in Fig.~\ref{Fig6} and Table~\ref{tableRR}.
The percolation threshold of the CP in sparse, undirected, uncorrelated complex networks of infinite size is known analytically to be $p_c = \langle\Delta\rangle/(\langle\Delta^2\rangle - \langle\Delta\rangle)$~\cite{Cohen00}.
In the cases of RRNs, $\langle\Delta\rangle=\Delta$ and $\langle\Delta^2\rangle=\Delta^2$ and so $p_c = 1/(\Delta-1)$ for the CP.
The percolation threshold of the CP in ERNs is $p_c = 1/\langle\Delta\rangle$ since $\langle\Delta^2\rangle=\langle\Delta\rangle^2+\langle\Delta\rangle$ in ERNs~\cite{Newman00}.
We confirmed that all $p_c$ values obtained numerically in this work are consistent with these analytic results within 0.05\%.
Besides, Table~\ref{tableRR} verifies that the BP with $m=1$ and $m=2$ has the same percolation threshold as the CP~\cite{Adler90}.

\section{Summary}

We presented a new efficient algorithm that applies to the BP model in nonregular networks within the Newman-Ziff algorithm.
With the new algorithm and the algorithms of Refs.~\cite{Choi19,Choi20}, we calculated the strength of the giant cluster $P_{\infty}(p)$ and its derivatives for the CP, DP, and BP in RRNs and ERNs.
Based on the results, we classified the percolation transitions. 
The DP with a small $k$ has the double transition, which is a second-order percolation transition followed by the discontinuous jump of $P_{\infty}$ at higher $p$.
The DP with a large $k$, CP, and BP with a small $m$ shows the second-order percolation transition.
The BP with $m\geq3$ has the first-order percolation transition.
The third-order percolation transition occurs in the BP of $m=2$ in RRNs.

The DP and BP are basic models for various real phenomena in complex systems such as neuronal activity, jamming transition, opinion formation, and diffusion of innovations.
Therefore, the results of this work, especially, the detailed analysis of the percolation transitions and the discovery of third-order phase transition would help to further understand dynamics and critical transformations in complex systems.

\section*{Acknowledgments}
This work was supported by the National Research Foundation of Korea(NRF) grant funded by the Korea government(MSIT) (No. 2021R1F1A1052117).

\bibliographystyle{elsarticle-num}
\bibliography{perc}

\end{document}